# Interplay between local structure and electronic properties on CuO under pressure.

V. Cuartero*[,1,2,3], V. Monteseguro[4,2], A. Otero-de-la-Roza[5], M. El Idrissi[2,6], O. Mathon[2], T. Shinmei[7], T. Irifune[7] and A. Sanson[6]

[1] Centro Universitario de la Defensa de Zaragoza. Ctra. Huesca s/n, 50090 Zaragoza (Spain).
[2] ESRF-The European Synchrotron, 71 Avenue des Martyrs, Grenoble (France).
[3] Instituto de Nanociencia y Materiales de Aragón (INMA), Departamento de Física de la Materia Condensada, CSIC-Universidad de Zaragoza. C/ Pedro Cerbuna 12, E-50009 Zaragoza (Spain).
[4] Departamento de Física Aplicada-ICMUV, Universidad de Valencia, MALTA Consolider Team, Edificio de Investigación, C/Dr. Moliner 50, 46100 Burjassot, Valencia (Spain).
[5] Departamento de Química Física y Analítica and MALTA-Consolider Team, Facultad de Química, Universidad de Oviedo, 33006 Oviedo, (Spain).
[6] Department of Physics and Astronomy - University of Padova (Italy).
[7] Geodynamics Research Center (GRC), Ehime University, 2-5 Bunkyo-cho, Matsuyama 790-8577 (Japan).

***Corresponding Author: vcuartero@unizar.es**

KEYWORDS. Strongly correlated electron systems; X-ray Absorption Spectroscopy; high pressure; *ab-initio* calculations.

**ABSTRACT**. The electronic and local structural properties of CuO under pressure have been investigated by means of X-ray absorption spectroscopy (XAS) at Cu K edge and *ab-initio* calculations, up to 17 GPa. The crystal structure of CuO consists of Cu motifs within $CuO_4$ square planar units and two elongated apical Cu-O bonds. The $CuO_4$ square planar units are stable in the studied pressure range, with Cu-O distances that are approximately constant up to 5 GPa, and then decrease slightly up to 17 GPa. In contrast, the elongated Cu-O apical distances decrease continuously with pressure in the studied range. An anomalous increase of the mean square relative displacement (EXAFS Debye Waller, $\sigma^2$) of the elongated Cu-O path is observed from 5 GPa up to 13 GPa, when a drastic reduction takes place in $\sigma^2$. This is interpreted in terms of local dynamic disorder along the apical Cu-O path. At higher pressures (P>13 GPa), the local structure of $Cu^{2+}$ changes from a 4-fold square planar to a 4+2 Jahn-Teller distorted octahedral ion. We interpret these results in terms of the tendency of the $Cu^{2+}$ ion to form favorable interactions with the apical O atoms. Also, the decrease in Cu-O apical distance caused by compression softens the normal mode associated with the out-of-plane Cu movement. CuO is predicted to have an anomalous rise in permittivity with pressure as well as modest piezoelectricity in the 5-13 GPa pressure range. In addition, the near edge features in our XAS experiment show a discontinuity and a change of tendency at 5 GPa. For P < 5 GPa the evolution of the edge shoulder is ascribed to purely electronic effects which also affect the charge transfer integral. This is linked to a charge migration from the Cu to O, but also to an increase of the energy band gap, which show a change of tendency occurring also at 5 GPa.

## INTRODUCTION.

CuO has been the subject of extensive investigation as understanding the nature of Cu-O bonds is key for the comprehension of cuprate-based high $T_c$ superconductors. More recently, CuO itself has seen renewed interest due to the discovery of multiferroicity (MF) at relatively high temperature $T_N$ = 230 K and ambient pressure (1). CuO presents type II MF, where ferroelectricity is magnetically driven by an antiferromagnetic (AFM) spiral ground state. These findings motivated theoretical (2, 3) and experimental studies considering the application of high pressure to promote a room temperature MF state in this material (4). These studies predicted stable MF at room temperature (RT) in the P~20-40 GPa range due to large super-exchange correlations in an incommensurate AFM ground state (3). Subsequently, Jana and coworkers (4) found an anomaly in the dielectric constant and a drop in DC resistance by three orders of magnitude at ~ 4 GPa and RT, that was proposed to be correlated with strong dynamic O-ion displacements along the *b*-axis in monoclinic CuO. In parallel, neutron diffraction experiments under pressure combined with Monte Carlo simulations found that $T_N$ is far from RT at 38 GPa. (5).

The crystallographic structure of CuO is monoclinic with *C2/c* space group. It consists of corner- and edge-sharing square-planar $CuO_4$ units, which form $(\text{-Cu-O-})_\infty$ zigzag chains along the [10-1] and [101] directions of the unit cell. Cu ions are coordinated by four O ions with interatomic Cu-O distances $R_1$ = 1.95 Å and $R_2$ = 1.96 Å, and there are two apical O ions at an elongated distance $R_3$ = 2.78 Å (6)(see figure 1). This local structure can be seen as a highly distorted 4+2 octahedron. The AFM Cu-O-Cu magnetic ordering is stabilized through super-exchange interactions (7) via the $b_{1g}$ ($d_{x^2-y^2}$) orbital, which is singly occupied under the $D_{4h}$ crystal field (8). The strongest $J_z$ superexchange interactions are mediated by Cu-O-Cu interactions, and occur along Cu-O-Cu chains characterized by the $\varphi_{[10\text{-}1]}$ angle, represented in figure 1.

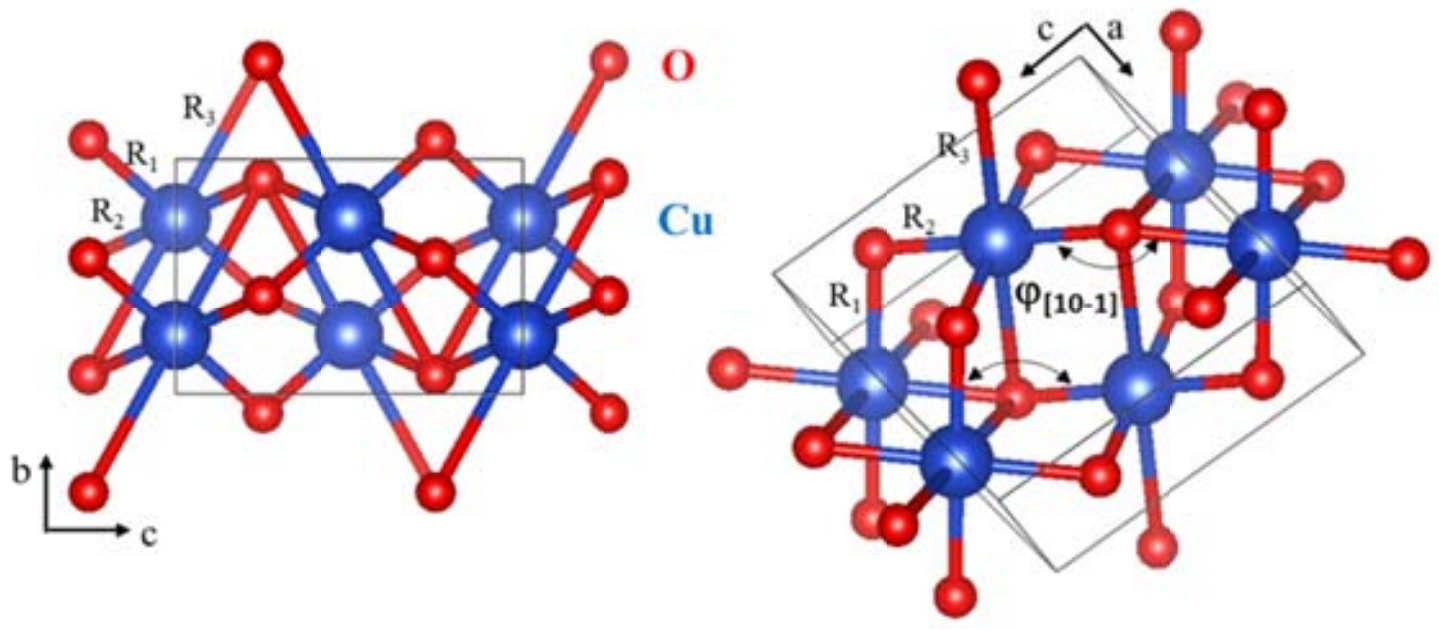

**Figure 1**. CuO monoclinic unit cell projected on the *bc* plane (left) and on an orientation where the $\varphi_{[10\text{-}1]}$ Cu-O-Cu angle is marked (right). Cu atoms are in blue and O atoms are in red. The average between $R_1$ and $R_2$ correspond to the $\text{Cu-O}_{short}$ distances and $R_3$ stands for the $\text{Cu-O}_{long}$ distance. The unit cell is marked by black lines.

X-ray (neutron) diffraction techniques provide information on the average long-range ordered crystallographic (magnetic) structures. Diffraction experiments have found no pressure-induced structural phase transitions in CuO up to 38 GPa using different pressure transmitting media (PTM) (4, 5, 9). The experiments also showed that the compression of the unit cell is anisotropic, and the *a*-axis expands up to ~ 13 GPa, and then decreases at higher pressures. In addition, Jana and coworkers observed subtle anomalies such a small kink in the normalized pressure-Eulerian strain curve at 3.4 GPa, ascribed to an electronic transition when there is no evidence of a structural transition (4). In contrast, Kozlenko and coworkers observed a smooth evolution of the lattice parameters and *y*-Oxygen coordinate along the whole pressure range, with a stable magnetic structure (5).

There is a delicate balance between lattice, electronic and magnetic properties in CuO. The compression of the Cu-O bonds plays a crucial role in the behavior of this material under pressure. In this article, we study the evolution of the Cu-O bonds and the electronic properties of CuO under pressure by means of X-ray Absorption Spectroscopy (XAS) and *ab-initio* calculations. XAS is a local probe that allows a chemical selective investigation of the electronic and local structural properties of the absorbing atom. In particular, we present extended X-ray absorption fine structure (EXAFS) measurements on CuO at Cu K edge under pressure, in order to get information on the evolution of the local structure around Cu ions. We obtained the evolution of the Cu-O bond distances and the EXAFS Debye Waller factors ($\sigma^2$) of the backscatterer with respect to the absorber atom, which parameterize the effects of structural and vibrational disorder (10), up to 17 GPa at RT. Moreover, the X-ray Absorption Near Edge Structure (XANES) region of the spectra sheds light on the pressure induced changes in the electronic density of states above the Fermi level. Furthermore, *ab-initio* calculations were performed within the framework of density functional theory, in order to corroborate and interpret the experimental results obtained by XAS.

## METHODS.

XAS measurements at Cu K edge were undertaken in CuO polycrystalline commercial powders (Aldrich, +99.9% purity) at high pressure up to 17 GPa. The experiment was performed at BM23 beamline at the ESRF (Grenoble, France) (11), equipped with a double crystal Si (111) monochromator and Kirkpatrick−Baez mirrors to focus the monochromatic x-ray beam down to 5 x 5 $\mu m^2$, with a Pt coating. The mirrors were set at an angle of 6 mrad to reject higher order harmonics. The powdered sample was loaded in a nano-polycrystalline diamond anvil cell (to avoid glitches from the anvils on the XAS data) (12), with ruby chips as pressure markers. Ne gas was used as pressure transmitting medium in order to be fully hydrostatic along the whole pressure range. A Cu foil was measured before every data point to ensure energy calibration.

The EXAFS spectra, $\chi(k)$, were obtained after removing the signal background. The Fourier Transform (FT) curves of the k weighted EXAFS signals were obtained for the k range 1.5 – 14.3 Å$^{-1}$ using a sinus window, with the Athena software of Demeter package (13). The EXAFS structural analysis was performed on the FT space by fitting these spectra with Ifeffit using the theoretical phases and amplitudes calculated by FEFF – 8 code (14), considering the atomic positions at RT and ambient pressure from reference (6).

XANES spectra were normalized to unity edge jump by using also Athena software from the Demeter package (13). Monoelectronic calculations of the XANES at Cu K edge have been carried out with FDMNES code (15, 16). FDMNES code is used to calculate XANES under the Green formalism in the muffin-tin approach, but it also provides the option to use the finite difference method to have a calculation with a free potential. The potentials are obtained by a self-consistent calculation until convergence. The cluster geometry was fixed to the structural determination (5, 6), considering the local spin-density approximation (LSDA) and including Hubbard correction (U) on the potential calculation.

*Ab-initio* total-energy calculations at zero temperature have been performed within the density functional theory (DFT) (17), by means of the VASP package (18-20), using the pseudopotential method and the projector augmented-wave scheme (PAW) (21). The calculations were carried out with a unit cell containing 8 atoms. For copper, 11 valence electrons were used ($3d^{10}4s^1$) whereas 6 valence electrons ($2s^22p^4$) were used for oxygen. Highly converged results were achieved by extending the set of plane waves up to a kinetic energy cutoff of 550 eV. In order to provide a reliable description of the effects of electronic correlation, the calculations were performed using the GGA + U formalism with the Dudarev's approach (22). The effective on site Coulomb and exchange parameters were set to U = 9 eV and J = 1 eV, yielding reliable results for the magnetic moments and the cell parameters as compared to experiments (5). A dense special k-points sampling (7 9 6) for the Brillouin Zone (BZ) integration was performed in order to obtain very well converged energies and forces. At each selected volume, the structures were fully relaxed to their equilibrium configuration through the calculation of the forces and the stress tensor. In the relaxed configurations, the forces on the atoms are less than 0.006 eV/Å and the deviation of the stress tensor from a diagonal hydrostatic form is less than 0.1 GPa.

The vibrational properties and the piezoelectric tensor were calculated by using the density functional perturbation theory (DFPT). Lattice-dynamics calculations were performed at the zone center (Γ point) of the BZ with a 3x3x3 supercell. The construction of the dynamical matrix at the Γ point of the BZ involves separate calculations of the forces in which a fixed displacement from the equilibrium configuration of the atoms within the primitive cell is considered. The number of such independent displacements in the analyzed structures is reduced due to the crystal symmetry. Diagonalization of the dynamical matrix provides the frequencies of the normal modes and it was realized by using PHONOPY software (23). Moreover, these calculations allow identifying the symmetry and eigenvectors of the vibrational modes in each structure at the Γ point. The piezoelectric tensor was determined including the ionic relaxation contributions to the elastic moduli.

## RESULTS AND DISCUSSION.

### A. Extended X-ray Absorption Fine Structure at Cu K edge.

CuO has a monoclinic structure, space group *C2/c*, with lattice parameters at ambient conditions a = 4.6837 Å, b = 3.423 Å, c = 5.129 Å, β=99.54º (6). The local structure around Cu atoms can be considered as 4 Cu-O short almost regular distances (Cu-$O_{short}$ = 1.956 Å) and 2 longer Cu-O distances at Cu-$O_{long}$ =2.784 Å. These are identified as $R_1$, $R_2$ (short) and $R_3$ (long) respectively in figure 1.

The Fourier Transformed EXAFS signals are plotted in figure 2 (a) for the whole pressure range and the corresponding selected EXAFS signals are shown in figure 2 (b). The low pressure spectrum agrees with monophasic CuO previously reported (24, 25). The first intense peak on the FT signal at ~ 1.5 Å corresponds to the 4 Cu-O shortest distances and it barely changes its position with increasing pressure, while its intensity is slightly reduced with compression. The following structures correspond to the contributions from Cu-$O_{long}$ (N=2) and Cu-Cu nearest neighbor paths. The analysis of the FT is performed by considering the starting model of the CuO structure at ambient pressure and RT (6). The fits are performed in R-space using a sinus window within the range R: [0.8, 3.2] Å, and six paths are included for all pressure points: [1] Cu-O (N=2) at 1.95 Å, [2] Cu-O (N=2) at 1.96 Å, [3] Cu-O (N=2) at 2.78 Å, [4] Cu-Cu (N=4) at 2.90 Å, [5] Cu-Cu (N=4) at 3.08 Å and [6] Cu-Cu (N=2) at 3.17 Å.

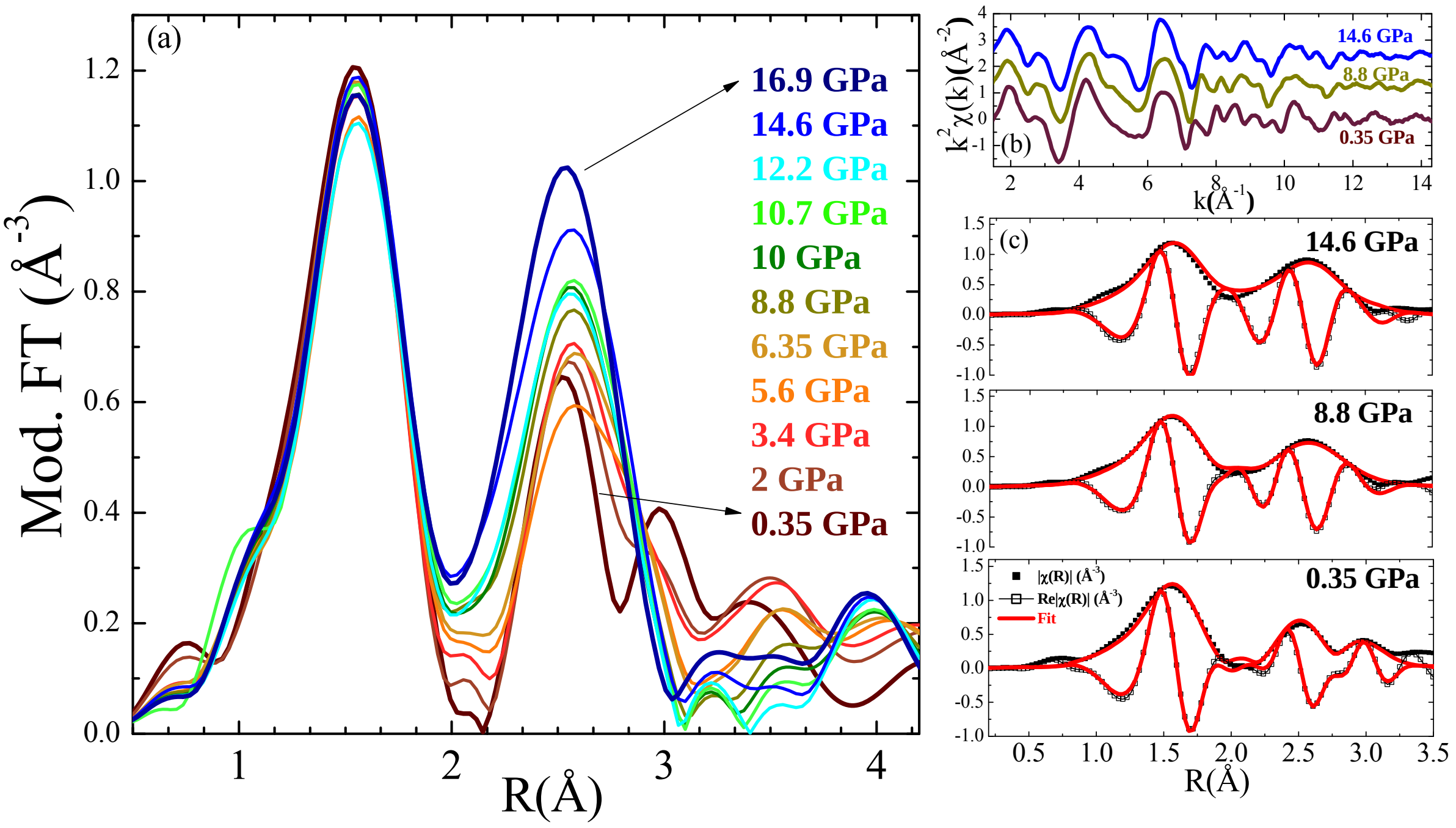

**Figure 2.** **(a)** Modulus of the Fourier Transformed EXAFS signals along the whole pressure range, considering a sinus window k: [1.5,14.3] Å$^{-1}$. **(b)** EXAFS signal weighted in $k^2$ for CuO for selected pressures. Spectra have been shifted vertically for the sake of clarity. **(c)** Modulus and real part of selected Fourier transformed EXAFS signals (points) and their corresponding fit (red lines) for selected pressures.

The calculated contributions from all paths are detailed in figure S1. The coordination numbers are fixed to their crystallographic values, and the amplitude reduction factor, $S_0^2$ and the energy shift parameter $\Delta E_0$ are fitted for the lowest pressure point (0.35 GPa) and then set to the obtained values for the rest of the fits (i.e. $S_0^2$= 0.78 and $\Delta E_0$ = 0.9 eV). We considered the same shift of the distances and Debye Waller (DW) $\sigma^2$ on the equatorial plane (paths [1] and [2], Cu – $O_{short}$, $delr_1$ and $\sigma^2{}_1$), and different ones for the apical longer path (path [3] Cu - $O_{long}$, $delr_2$ and $\sigma^2{}_2$). The Cu-Cu paths [5] and [6] are grouped and the same shift and $\sigma^2$ factor is used for both, as this combination provided the best fitting R-factors, as detailed Table S1. The fitted curves for three selected pressures are shown in figure 2 (c). The R-factors of the fit are below 0.01 in all cases. More fits have been tested using different k-weights and with a Gaussian window, giving equivalent results for the evolution of the fitting parameters with pressure. The results of the fits for the first shell are shown in figure 3, and the values obtained at ambient pressure are in agreement with previous EXAFS results (24). The error bars are obtained from the standard deviation of the different values from the average considering the results from fits with three different k-weights (k=1, 2 and 3). The evolution of the Cu-O distances with pressure, obtained by EXAFS and by the *Ab-initio* calculations, is plotted in figure 3 (a). The *ab-initio* calculated values (plotted by dotted lines) are in good qualitative agreement with the experiment, except for the between the distances in the low pressure range (P< 5 GPa). The correlations obtained by EXAFS analysis between the Cu-$O_{long}$ and the Cu-Cu [4] distances are quite important for the P< 5 GPa region. In order to overcome this issue, the Cu-Cu [4] shortest distances, which are the closest to the Cu-

$O_{long}$ distance, are fixed to their crystallographic values and the fit is repeated with the rest of the parameters set as free. The resulting Cu-$O_{long}$ distances are slightly higher than the ones obtained by the first analysis and are closer to the *ab-initio* calculated values. Then, the error bars on the Cu-$O_{long}$ distances obtained below 5 GPa are corrected accordingly, by considering the error introduced also by the second fit value. The lattice parameters were also calculated, and they are in accordance with the ones found in ref. (5) (not shown here).

Figure 3 (a) shows that the shortest Cu-O distances in the square planes are barely affected by pressure, being constant up to ~ 5 GPa within the experimental error bar. Above that pressure, the distance is slightly reduced, and the bond compressibility, defined as $k_i=-1/d_i(\partial d_i/\partial P)$ (26), is $3.6\cdot10^{-4}$ $GPa^{-1}$. It is worth mentioning that the same evolution of Cu-$O_{short}$ distance is found when the variation of $E_0$ parameter is left free on the EXAFS fitting, and also for the fit including just the first shell. The evolution with pressure of the longest Cu-O distances are quite different, as they are continuously reduced along the whole pressure range, with compressibility $6.7\cdot10^{-3}$ $GPa^{-1}$, that is 18 times higher than the short square planar ones, in agreement with ref. (5). $Cu^{2+}$ is a $d^9$ metal and, therefore, a strong Jahn-Teller effect prevents it from acquiring an octahedral geometry. The octahedral coordination in CuO at ambient pressure and temperature is severely distorted due to the lengthened apical bond (it is longer than the basal bonds in more than 0.5 Å) (8) and it can be considered as square planar. We calculate the evolution of the Jahn-Teller distortion parameter, defined as $\sigma_{JT} = [\sum_{i=1}^{6}(R_{Cu-O} - \langle R_{Cu-O} \rangle)^2]^{1/2}$, where $R_{Cu-O}$ are the distances of Cu-O distorted octahedra and $\langle R_{Cu-O} \rangle$ is the average distance (26). The values are shown in the inset of figure 3 (a). We observe a reduction of the distortion of 42% upon compression, from 0.62 Å at 0.35 GPa down to 0.36 Å at 16.9 GPa. The theoretical $\sigma_{JT}$ is also included in the plot, showing a continuous decrease with compression as well.

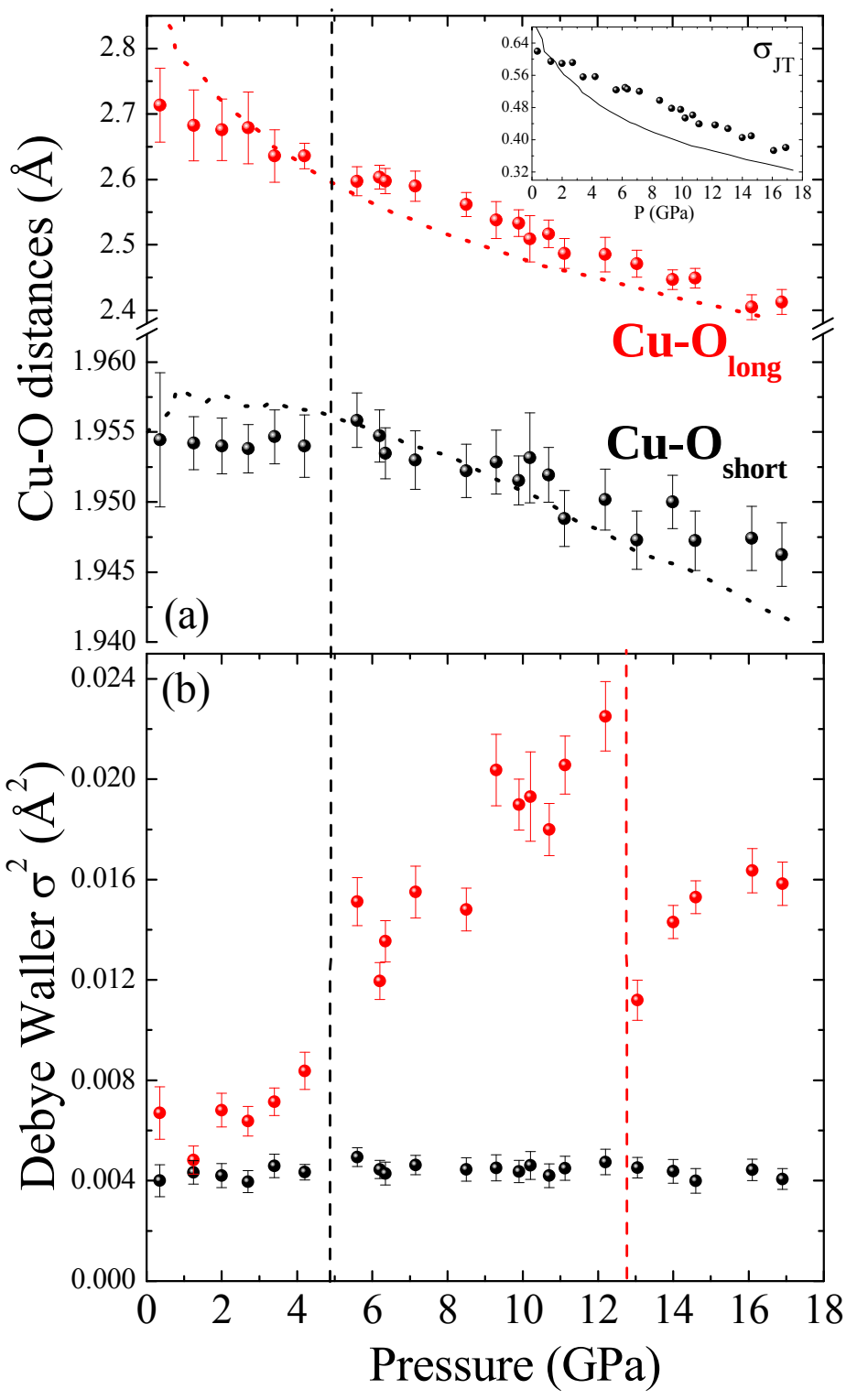

**Figure 3.** (a) Evolution of Cu-O distances with pressure. The Cu-$O_{short}$ distance corresponds to the average square planar ones (black points) and Cu-$O_{long}$ corresponds to the apical distances (red points). The corresponding evolution obtained by *ab-initio* calculations are plotted in dotted lines. (b) Evolution of the EXAFS DW factor $\sigma^2$ of the short (black) and long (red) distances with pressure.

The evolution with pressure of DW $\sigma^2$ factors corresponding to Cu-$O_{short}$ and Cu-$O_{long}$ paths is shown in figure 3 (b). The $\sigma^2$ factor of Cu-$O_{short}$ path is almost constant along the whole pressure range ($\sigma^2_{Cu\text{-}Oshort}$ ~ 0.004 $Å^{-2}$). Interestingly, we find a different behavior for the $\sigma^2$ corresponding to the Cu-$O_{long}$ apical distance, that we can divide in three regions: *(1)* from ambient pressure up to

~5 GPa, the DW barely changes upon compression, *(2)* from 5 GPa up to 13 GPa there is an anomalous significant increase of the DW factor, unexpected under compression, and *(3)* above 13 GPa, the DW drops down at a higher value than in region *(1)* and slightly increases, up to 17 GPa.

Moreover, from the EXAFS analysis we get the evolution of Cu-Cu distances and the DW factors, shown in figure S2. The Cu-Cu labeled as [4] above, are slightly reduced with pressure and the [5] and [6] Cu-Cu distances are reduced up to 5 GPa, and above that pressure they are slightly reduced up to 17 GPa. The compression of these distances observed by diffraction is more remarkable than the behavior observed here by EXAFS. This is most likely due to distance parameter correlation, as these distances are close. The DW factors (fig. S2 (b)) for [5] and [6] paths and change within the error bar (~0.008 $Å^{-2}$ ), while for path [4] it increases up to 5 GPa and then it slightly decreases upon compression.

**b. Interpretation of EXAFS results: local structure around $Cu^{2+}$ ions, vibrational and piezoelectric properties.**

EXAFS probes the instantaneous local structure around the absorbing atom (the interaction time of the XAS process is ~$10^{-14}$ s). The EXAFS DW factor assigned to the atomic distances contains two contributions: a thermal (or dynamic) one and a structural (or static) one. The first contribution is related to thermal disorder and the second one to local structural distortions, and they cannot be separated (10, 27). The rise of the DW factor of the Cu-$O_{long}$ distance in region *(2)*, starting at approximately 5 GPa, is ascribed to an increase in configurational disorder with pressure, either of static or dynamic origin, which means there is a wide distribution of Cu-$O_{long}$ distances in this material under compression in region *(2)*. Similar behavior has been observed in other systems (28, 29). In the context of the local structure of the $Cu^{2+}$ ions, the shapes of the local potential energy surface of the Cu atom give information on the particular distribution of Cu-O distances. This is shown in the inset of figure 4. The black parabola describes the equilibrium position of the Cu-$O_{long}$ distance at ambient conditions. The flattened (blue) potential energy curve corresponds to a situation where the ion has enough thermal energy to sample very different Cu-$O_{long}$ distances, but the energy minimum corresponds to the Cu atom at the center of the $CuO_4$ square, in agreement with a dynamic disorder. In contrast, the double-well (green) potential energy curve, in which the equilibrium position of the Cu is off the square plane, stands for a 4+1 square pyramidal coordination. In this later, the disorder is static and it originates from the thermal energy of the Cu being enough to create disorder between both off-center equilibrium positions. Our calculations do not agree with a double-well potential energy curve, but with the first situation (blue curve), which means there is a dynamic distribution of Cu-$O_{long}$ distances in this material under compression in region *(2)*.

According to diffraction experiments, there is no structural transition in CuO in this pressure range (5). This is in agreement with our *ab-initio* calculations, which predict that the $Cu^{2+}$ ion is always at the center of the $CuO_4$ motif, and Cu atoms are kept, on average, in a centro-symmetric crystallographic position.

Figure 4 shows the evolution with pressure of the average <Cu-$O_{long}$> distance (also shown in fig. 3), and the calculated maximum deviations Cu-$O_{long}+\delta$ and Cu-$O_{long}-\delta$, obtained from the associated DW factor (i.e. Cu-$O_{long}\pm \delta$ = <Cu-$O_{long}$> $\pm (\sigma^2-\sigma^2_{a.p.})^{1/2}$, where $\sigma^2_{a.p}$ is the DW factor at ambient pressure considering the error bar). The range spanned by these two distances is relatively small in region *(1)*, increases sharply at 5 GPa and remains high in region

*(2)*, and then it decreases again at 13 GPa and remains relatively small in region *(3)*.

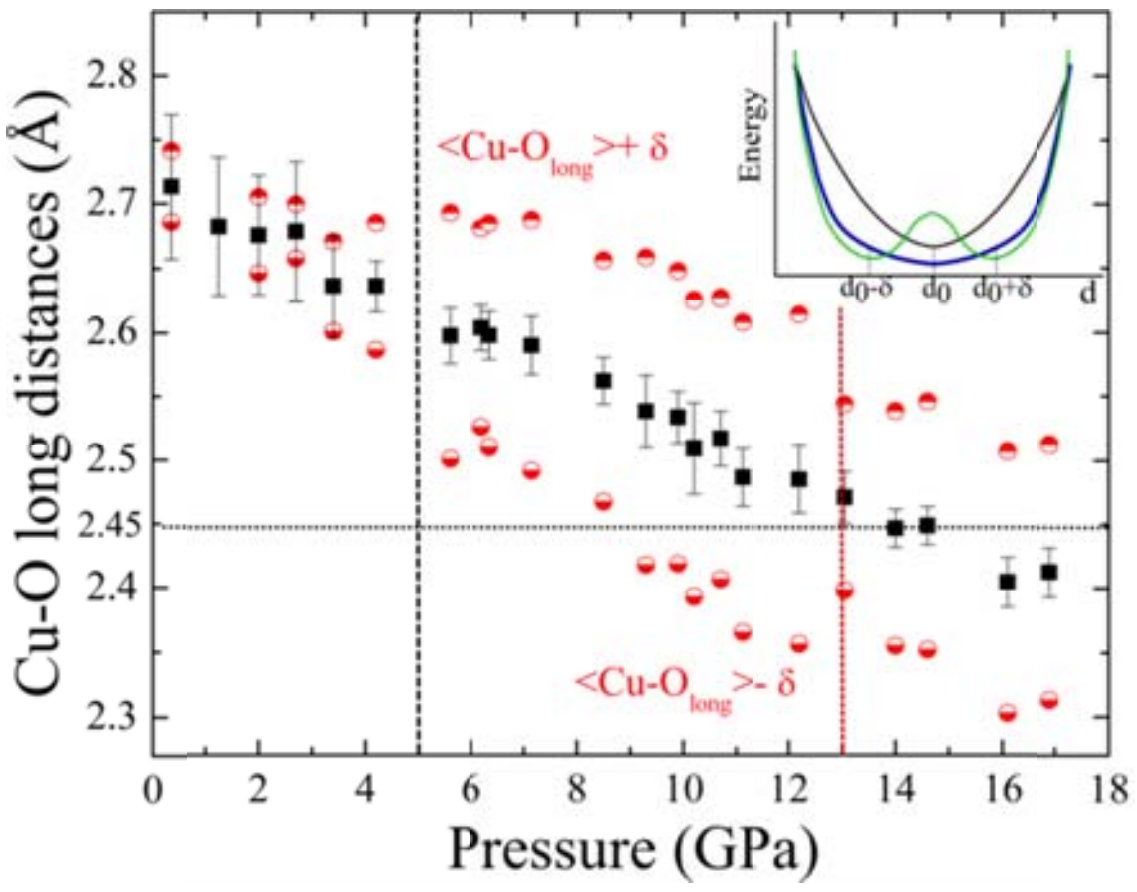

**Figure 4.** Average <Cu-$O_{long}$> distance obtained by EXAFS analysis (black squares) together with the calculated maximum deviation <Cu-$O_{long}$>±δ, obtained from <Cu-$O_{long}$> and the associated DW factor as a function of pressure. Inset: Energy profiles for the distribution of Cu-$O_{long}$ distances, where $d_0$ is the average distance. The black curve represents the profile at ambient pressure and temperature. The blue curve is the one obtained by *ab-initio* calculations and the green curve stands for a two different distances model.

The results from EXAFS measurements analysis can be rationalized by appealing to the chemistry of the $Cu^{2+}$ complexes. Being a $d^9$, an octahedral coordination for $Cu^{2+}$ would be unstable. It is known that the most common coordination of this ion is square-pyramidal (N=5), with 4+2 distorted-octahedron coordination also being common (8). In the cases when the square-pyramidal coordination happens, the apical Cu-O bond is longer than the basal ones by up to 0.5 Å (8). Figure 3 shows that the basal Cu-O bond distances barely change in the studied pressure range, with values between 1.945 Å and 1.955 Å and a relatively small Debye-Waller factor. Using Halcrow's value, this implies that a significant interaction between Cu and the apical O atom is expected at approximately 2.45 Å.

This approximate cutoff distance is shown in Figure 4 and compared to the EXAFS results, and it provides a key to interpreting the results. At zero pressure, the apical oxygen is more than 2.7 Å away from the Cu atom, too far to engage in a significant interaction (8). As pressure increases, the apical oxygen is forced near the $Cu^{2+}$ and this interaction increases in strength. Figure 4 shows that in region *(2)* the Cu-$O_{long}$ distance is still longer than 0.5 Å with respect to the shortest ones, but thermal motion of the Cu could bring the atom into a favorable interaction with one of the apical oxygen atoms (according to the blue curve of the energy profile in the inset). This instability definitely contributes to the increase of the EXAFS DW factor for the $CuO_{long}$ bond and, as we shall see below, it also causes the softening of the vibrational modes calculated in Γ point, in which the Cu moves out of the oxygen square plane. Of course, the increased interaction between Cu and one of the apical oxygens when the metal atom moves away from the centrosymmetric position comes at the cost of a destabilization of the opposite Cu-$O_{long}$ interaction. Our calculations and previous diffraction experiments (5) show that the Cu-$O_{long}$ interaction is never strong enough to stabilize a double-well potential energy curve (inset in Figure 4).

When the compression increases even further, the apical O is brought even closer to the Cu atom so that their distance reaches the 2.45 Å cutoff, with the Cu at the centrosymmetric position. The pressure at which this occurs coincides almost exactly with the transition to region *(3)* observed in EXAFS results (figure 4), at which point there is a substantial reduction in the DW factor and therefore in the disorder associated with the $CuO_{long}$ bond. The strength of the apical Cu-$O_{long}$ interaction becomes significant (8), and the coordination becomes a 4+2 distorted octahedron. We note that at 13 GPa there is also a pronounced change in the average pressure coefficient for *a* lattice parameter and the $\beta$ angle of the monoclinic structure (5), which becomes more distorted as observed in diffraction measurements. At even higher pressures (P>13 GPa), the Cu-$O_{long}$ distance is reduced even further and the 4+2 octahedral-like coordination is maintained, with enhanced disorder on the apical distance as the DW factor slightly increases (figure 3 (b)).

As previously mentioned, consistent with diffraction experiments, our calculations predict that, at all pressures, the Cu atoms always occupy the centrosymmetric site at the center of the $CuO_4$ motif. Examination of the calculated phonon density of states confirms that the calculated equilibrium geometries are true minima. By way of illustration, the phonon dispersion at 7.5 GPa is showed in the figure 5 (a). The phonon dispersion curves have been calculated along the high-symmetry k-point path (Γ-Y-E-A-B-Γ-D-Z-Γ) in the first BZ.

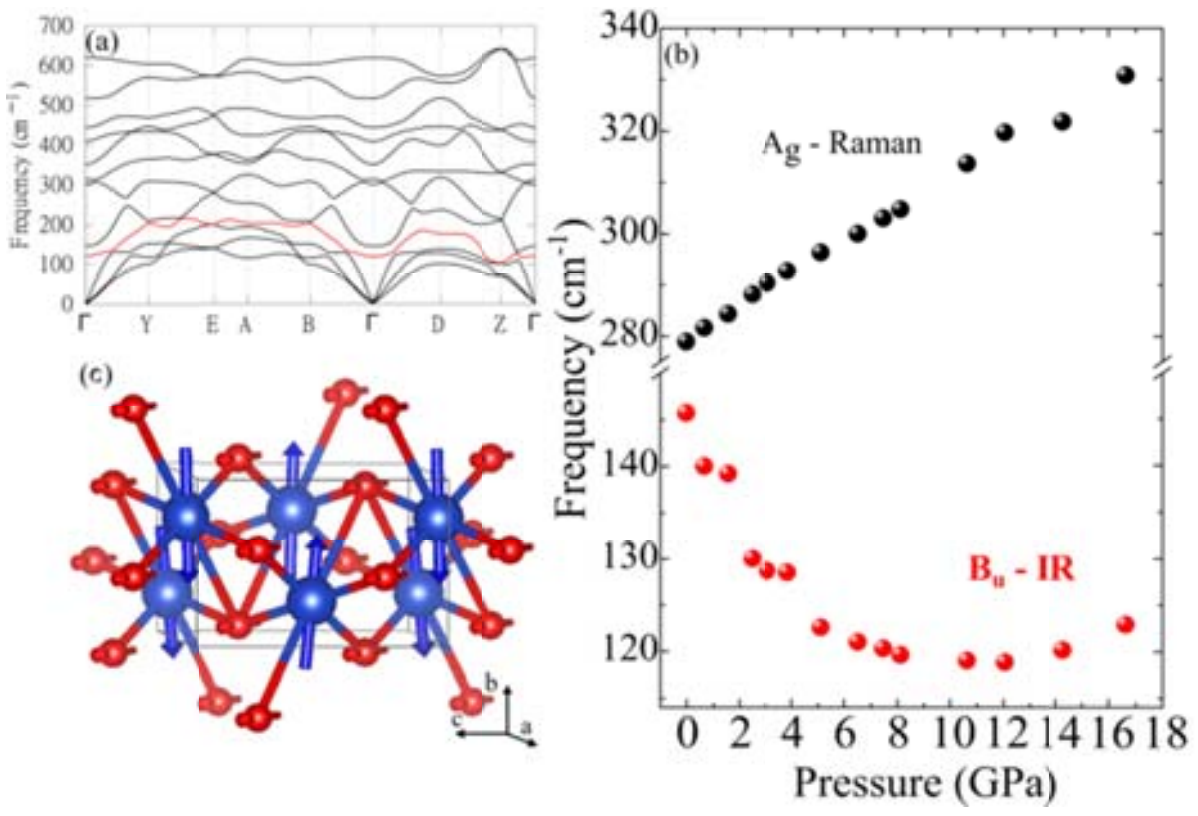

**Figure 5 (a)** Phonon dispersion curves calculated along a high-symmetry k-point path in the first Brillouin Zone at 7.5 GPa. The optical band corresponding to the $B_u$ (IR) mode at Γ is shown in red. **(b)** Theoretical pressure dependence of the Raman $A_g$ mode and the $B_u$ IR mode of CuO at zone center **(c)** Scheme of $B_u$-IR vibrational mode on the CuO unit cell.

Examination of the vibrational frequencies at the BZ center (the Γ point) can be used to probe the curvature of the potential energy curve of $Cu^{2+}$ ion as a function of pressure. Figure 5(b) shows the calculated evolution under compression of the $A_g$-Raman- and $B_u$-IR-active modes of CuO with main vibrational components along *b* axis (30). We are interested in these modes since the Cu-$O_{long}$ distances have their maximum projection along *b* axis. It is important to remark that these modes were calculated in the zone center. However, out of Γ point, there may be more phonons that have a significant influence in the vibration along *b* axis. $A_g$-Raman mode increases its vibrational frequency with compression, and the calculated pressure dependence is in quite good agreement with the Raman measurements from refs. (4) and (31). We focus our attention on the $B_u$ -IR mode (in red in figure 5 (a)), which is the first (lowest-energy) optical mode at zero pressure. Figure 5 (c) depicts the atomic displacements associated with this mode at the BZ center. This $B_u$ mode corresponds almost exactly to the motion in which the $Cu^{2+}$ ion is displaced perpendicularly away from the $CuO_4$ square plane motif (~15° deviation). Figure 5 (b) gives the pressure evolution of this $B_u$ mode. This mode softens under increasing pressure up to approximately 5 GPa, then remains constant at a relatively low frequency, and then the frequency increases at around 13 GPa. This coincides with the three pressure regions in figure 4 and it is consistent with our interpretation of the EXAFS results and our discussion regarding the shape of the $CuO_{long}$ potential energy curve. In region *(1)*, the Cu-$O_{long}$ interaction is too weak (relatively high frequency of the $B_u$ mode). Then, in region *(2)*, the strength of the interaction increases and the proximity of the apical O atom facilitates the motion in which the Cu atom breaks away from the centrosymmetric position (lower frequency). However, this effect is not enough to create an off-center energy minimum (the frequency is still positive). Lastly, in region *(3)* the apical O atom is even closer and the Cu atom has less tendency to escape from the centrosymmetric position, and therefore the $B_u$ frequency increases again. Figure 5 (a) shows that the examined $B_u$ mode has an even lower frequency away from the BZ center (red line), suggesting favorable interactions between local dipoles in adjacent cells. However, this frequency is never zero, and our calculations predict that the centrosymmetric position is always lowest in energy.

We also note that the distinction between the one-minimum and two-minima situations in CuO has important implications regarding the macroscopic properties of this material, as observed in other systems (32). Since the center of the $CuO_4$ motif is a centrosymmetric site and a displacement of the $Cu^{2+}$ cation creates a local dipole, we expect this material to be a candidate for ferroelectricity if it has a double-well potential energy curve, because it would be able to sustain a spontaneous polarization. In contrast, if a wide distribution of Cu-$O_{long}$ distances is dominant (wide single-minimum blue energy curve in figure 4), then we could expect a pressure-induced rise in the permittivity of the material but neither a spontaneous polarization nor ferroelectricity. This interpretation could also explain the anomalous behavior of the permittivity under applied pressure (4).

Finally, we calculated the piezoelectric tensor using finite differences at three pressures in each of the three regions (0, 7.5 GPa, and 14 GPa). Interestingly, the piezoelectric tensor is essentially zero at zero pressure, with $e_{ijk}$ values in the order of $10^{-3}$ C/m$^2$. In contrast, the piezoelectric tensor is non-zero for the longitudinal strains at 7.5 GPa. The electric displacement field induced by a longitudinal strain in the x direction is given by the piezoelectric constants $e_{111}$ = 0.16 C/m$^2$, $e_{211}$ = 0.068 C/m$^2$, and $e_{311}$ = 0.12 C/m$^2$. Interestingly, the displacement field induced by a longitudinal strain in the y direction ($e_{122}$ = 0.16 C/m$^2$, $e_{222}$ = 0.066 C/m$^2$, and $e_{322}$ = 0.12 C/m$^2$) and in the z direction ($e_{133}$ = 0.16 C/m$^2$, $e_{233}$ = 0.074 C/m$^2$, and $e_{333}$ = 0.12 C/m$^2$) are very similar, and the displacement induced by shear strains is essentially zero. The piezoelectric tensor decreases again at 14 GPa (region *(3)*), with $e_{111}$ = 0.045 C/m$^2$, $e_{211}$ = 0.067 C/m$^2$, and $e_{311}$ = 0.026 C/m$^2$. As in the previous case, the displacements are equivalent regardless of the direction of the longitudinal strain and essentially zero for shear strain.

## C. Electronic properties.

The Cu K-edge XANES spectra at representative pressures are shown in figure 6 (a). The spectra show two clear features at low energy, the first in the pre-edge region (A, at 8978 eV at ambient pressure) and the second at the edge (B, at 8986 eV at ambient pressure). The effects of pressure can be clearly seen in shoulder B, which is shifted towards higher energies, and they are more subtle in the small pre-edge structure A.

The weak A-peak on Cu K edge XANES in CuO is normally ascribed to quadrupole transitions to empty *d* states (33-35). In particular, the quadrupolar transitions 1*s* → $d_{x^2-y^2}$ have been identified as the most important contribution to the A pre-peak, as the lowest unoccupied states in CuO at the Cu site have $b_{1g}$ ($d_{x^2-y^2}$) character (34, 36). These orbitals lie within the $CuO_4$ square plane. Here we note a very subtle shift of the A pre-peak (not shown) towards lower energies on the previously described region *(2)* (from 5 up to 13 GPa). This can be linked to the rearrangement of the *d*-states ($e_g$ and $t_{2g}$), that become closer under compression, due to the slight shortening of Cu-$O_{short}$ distances in the square plane and to the evolution towards a 4+2 fold-like coordination.

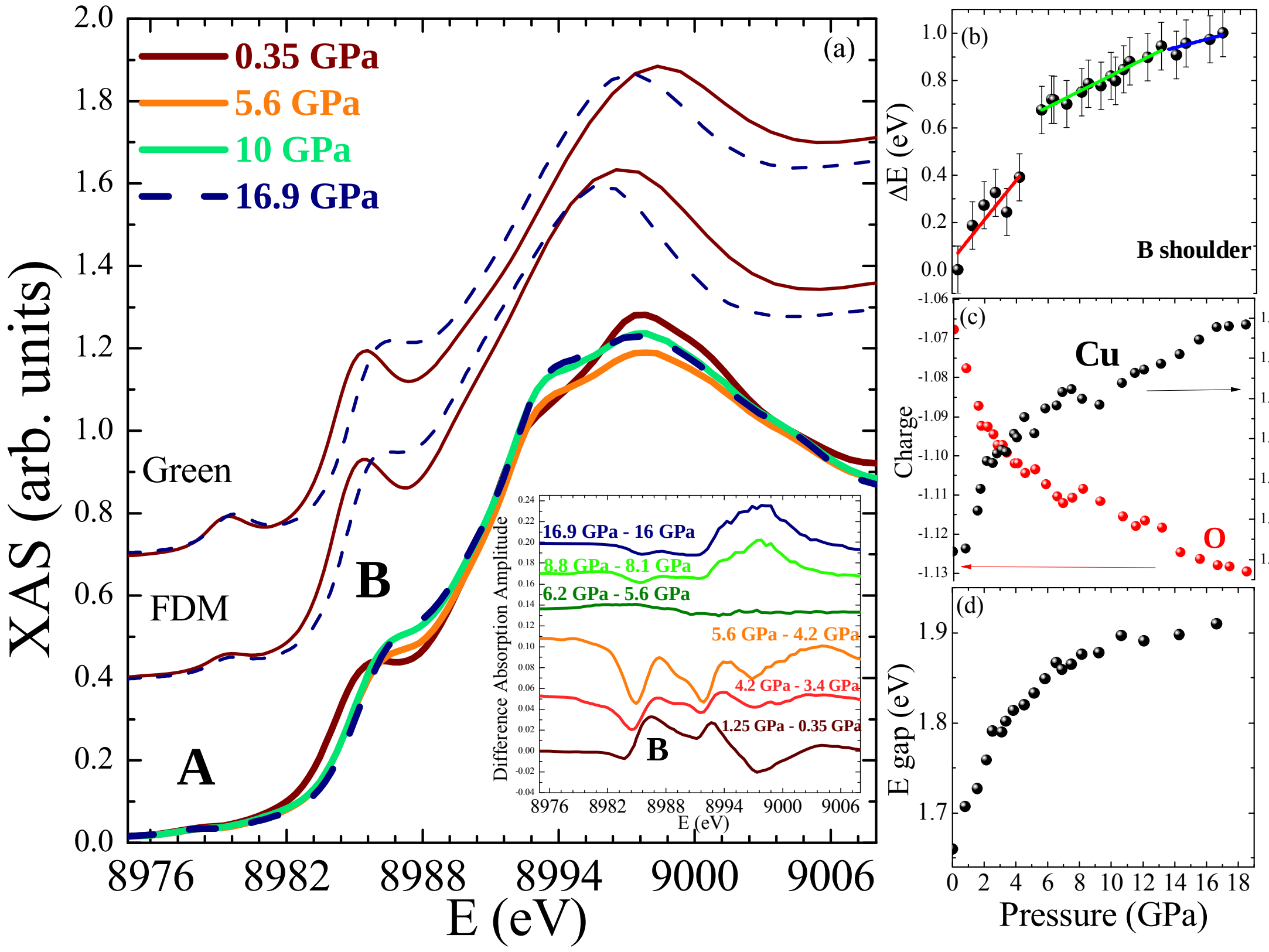

**Figure 6**. **(a)** XANES at different representative pressure points, from 0.35 up to 16.9 GPa (bottom spectra) and calculated XANES at ambient pressure and 17 GPa obtained by two methods: (1) Green formalism and muffin-tin potential and (2) Finite Difference Method (FDM). Inset: Difference between absorption spectra taken at successive pressures. Evolution with pressure of **(b)** the energy shift of the mean position of B structure, **(c)** the Bader charges of O, and Cu ions and **(d)** the energy band gap.

The B shoulder energy mean position changes with compression as can be seen in figure 6 (a) and (b). The origin of the B feature is quite controversial and there have been many attempts in the literature to calculate it by means of multiple scattering methods (25, 34, 36-38). Some authors claim that the position of B originates from multiple scattering effects of the photoelectron within the molecular cage defined by the first shell, which supports a one-electron interpretation of this shoulder (37). Other theoretical studies say that it has a mixed origin: in part it is one electron and in part it is caused by many-body shake-down processes (34), as earlier claimed by several authors (39-41), and more widely accepted nowadays (42, 43). In the latest case, the B shoulder structure is ascribed to a group of dipole-allowed transitions involving mainly $1s \rightarrow 4p_\pi$ excitations, simultaneous to the ligand-to-metal charge transfer (CT) process. In other words, the B spectral feature originates from dipolar electronic transitions from the 1*s* core level up to $4p_\pi$ orbitals (*p* apical orbitals, in terms of the molecular orbital notation), accompanied by a shakedown process through CT from the 2*p* oxygen ligands (L) to the Cu 3*d* hole, which enhances the screening of the core hole leading to a final excited state $|\Psi> = \beta|1s^1\ 3d^{10}\ \underline{L}\ 4p_\pi^{\ 1}> + \alpha|1s^1\ 3d^9\ 4p_\pi^{\ 1}>$, where $\underline{L}$ represents the ligand hole.

In order to further validate our XAS measurements, theoretical calculations have been performed to obtain the simulated XANES under two different approaches. The first uses the Green formalism on a muffin-tin potential and the second uses a free shape potential by means of the finite difference method (44). In both cases, the Hubbard correction is included in the calculations (LSDA+U), considering U= 9.79 eV (45). The cluster geometry is fixed to the structural determination, as found in reference (6) for the ambient pressure phase and in ref. (5) for the high pressure phase (15 GPa). The cluster size is 6 Å (93 atoms), as no changes are observed on the calculations for higher radius values. The convolution of the spectra is performed with a Lorentzian with FWHM of 1.5 eV to simulate 1 *s*-hole lifetime broadening and a Gaussian to simulate the experimental resolution.

The figure 6 (a) shows the calculated spectra using the two methods. The qualitative agreement with the experiment is quite acceptable in both cases, and the relative energy positions of the spectral features are coincident, which validates the experimental data. Interestingly, the evolution of B structure towards higher energies is well reproduced. The pre-peak A is less intense on the FDM calculation, and it matches better the experimental intensity ratio with the other features. Even though the edge features on Cu K edge XAS of CuO are ascribed both to multiple scattering (MS) and shake-down processes through charge transfer, MS calculations usually give qualitative agreement, as the ligand field and the geometrical structure can also modify the 1s→4p transition probabilities.

The experimental XANES under pressure shows that the B-shoulder structure is progressively smoothed (figure 6 (a)) and its position is slightly but clearly shifted to higher energies with compression, as represented by its mean position in figure 6 (b). This has been obtained by removing the background and by fitting the peak to a Lorentzian curve to extract the central position. We observe in figure 6 (b) a linear and continuous shift up to a discontinuity at 5 GPa. As observed for Cu-O distances and DW factors evolution under compression, the same three different regions can be distinguished on the trend of B shoulder with energy under pressure: *(1)* from ambient pressure up to 5 GPa, with ΔE = 0.4 eV, *(2)* from 6 GPa up to 13 GPa, with ΔE = 0.25 eV, *(3)* from 13 GPa and on, with a smaller shift of the mean position. On the transition from region *(1)* to *(2)*, we observe a jump at 5 GPa of 0.3 eV. To check the presence of small changes on the different pressure regions, the spectra obtained at successive pressure were subtracted from each other. The difference is plotted in the inset of fig. 6(a). The largest difference at the edge position corresponds to the 5.6 GPa – 4.2 GPa spectra, just at the transition between regions *(1)* and *(2)*. Besides, the spectral dependence of the differential signal around B structure is similar on region *(1)* and changes in region *(2)*, where the structures almost disappear.

The B shoulder position depends significantly on the local symmetry in the case of $Cu^{2+}$ compounds (i.e. $La_2CuO_4$, $Nd_2CuO_4$ (41)). However, our EXAFS results (figure 3 (a)) demonstrate that, in this case, the B shoulder shift below 5 GPa cannot be linked to any local structural variation of the Cu-$O_{short}$ distances in the square plane, as these are not changing with compression in this pressure range. Therefore, the shift towards higher energies in region *(1)* must be ascribed to an electronic contribution coming from the absorber under compression. In order to shed light on this experimental observation, we calculated the evolution of the Cu and O ionic charges using Bader charge analysis (46), and the evolution of band gap (fig. 6 (c) and (d)). CuO is a charge transfer (CT) semiconductor (47) because its energy bandgap is proportional to Δ, the charge-fluctuation transfer energy $3d^9 \rightarrow 3d^{10}\underline{L}$. This is found for those systems where *U*, which includes exchange interactions and *d-d* Coulomb interaction, is larger than the CT energy, Δ. Then, an increase of Δ entails an increase of $E_{gap}$. Figure 6 (c) shows the evolution under pressure of the Bader charge of Cu and O ions obtained from *ab-initio* calculations. There is a subtle but continuous migration of charge from the Cu ions towards O ions, which is consistent with an increase in the crystal field with pressure, with the *p-d* bands getting closer in energy, and with an increase in the charge transfer energy. The subtle increase in the ionicity of the system is more pronounced below 5 GPa, coinciding with the region *(1)* where the B shoulder mean position has a steeper dependence with pressure.

The evolution with pressure of the energy band gap is shown in figure 6 (d). As expected for a CT semiconductor, there is an increase of the energy band gap which behaves similarly to the metal-to-ligand charge transfer under compression. Moreover, 3 pressure regions can be distinguished in the calculated parameters in figures 6 (c) and (d) to point out the parallelism between the charge transfer, the energy band gap and the B shoulder energy-position dependence with pressure. In all cases there is a change of slope around 5 GPa, a smooth increase up to ~11-13 GPa and smaller variation up to 17 GPa. The change of tendency and the jump of 0.3 eV at 5 GPa, coincides with the onset of Cu-$O_{short}$ contraction and the beginning of the increase of dynamical disorder on Cu-$O_{long}$ (region *(2)*). Therefore, region *(1)* evolution can be explained in terms of metal to ligand charge transfer effects, which reduce the covalency and increases charge delocalization. Furthermore, the very smooth shift of 0.25 eV up to 13 GPa (region *(2)*) can be also correlated to the continuous compression of the $CuO_6$ volume under pressure. Again, above ~13 GPa either ΔE or the charge (or $E_{gap}$) barely change, which supports the hypothesis of the stabilization of the local structure.

Interestingly, the angle $\varphi_{[10\text{-}1]}$ along the Cu-O-Cu chain (marked in figure 1**)**, responsible for the strongest $J_z$ superexchange interaction (2), shows a change of trend around 5 GPa as well, as shown in figure 7. The angle obtained from our *ab-initio* calculations is compared with the experimental values found by diffraction by Kozlenko *et al.* (5), showing a similar trend. In region *(1)*, below 5 GPa, the angle rapidly increases with a rate of 4 °/GPa while in the region *(2)* (P> 5 GPa) this angle barely increases with a rate of 1 °/GPa. These trends in both regions match very well with those obtained by X-ray diffraction (2 °/GPa and 1 °/GPa) and neutron diffraction (4 °/GPa and 1°/GPa.). This behavior is consistent with the calculated pressure dependence of the metal-to-ligand CT and of the energy band gap, remarking the strong coupling between the magneto-electric properties in CuO.

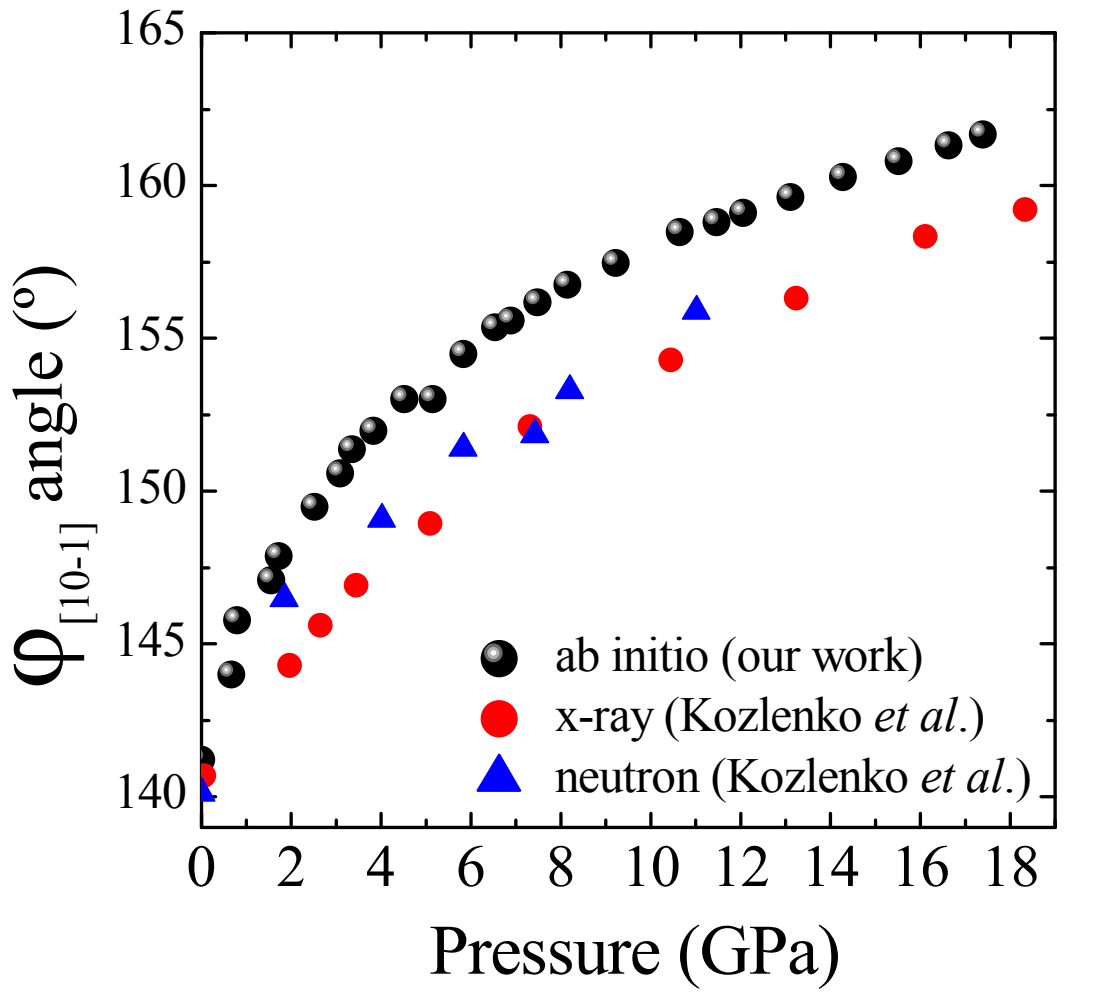

**Figure 7.** Pressure dependence of the $\varphi_{[10\text{-}1]}$ angle obtained by *ab-initio* calculations (black points) and by diffraction experiments from reference (5).

## CONCLUSIONS.

The local structural and electronic properties of CuO under pressure were investigated by X-ray absorption spectroscopy and *ab-initio* calculations. The $CuO_4$ square planar units are found to be stable under compression up to 17 GPa, while the apical Cu-$O_{long}$ distance is continuously decreasing under pressure. The absolute value of the compressibility of Cu-O bonds is slightly smaller than the one found in diffraction measurements, although the relative difference between the shorter and longer Cu-O bonds compressibility is similar (5).

The EXAFS Debye Waller factor, $\sigma^2$, does not change with pressure for Cu-$O_{short}$ bonds, while it shows a continuous increase for Cu-$O_{long}$ bonds (figure 3 (b)) from 5 GPa up to 13 GPa, when a sudden drop takes place. This anomalous rise under compression suggests an increase of configurational disorder of the Cu-$O_{long}$ bonds. This means that there is a distribution of different Cu-$O_{long}$ distances and as a result there are dynamically distorted Cu-O local environments. At 13 GPa, the DW factor $\sigma^2$ is reduced by more than 30% of its value due to the onset of a distorted octahedral local structure, which coincides with a pronounced change in the average pressure coefficient for $a$ lattice parameter and the $\beta$ angle of the monoclinic structure (5). According to our calculations and experimental results, the equilibrium position of the Cu ion is the center of the basal plane and, in average, in a centrosymmetric position. We interpret the anomalous behavior of the DW factor of the Cu-$O_{long}$ bonds under compression in terms of the ability of the Cu atom to interact with the apical O atoms. Because the Cu is most stable in a centrosymmetric site, the system cannot maintain polarization, and the presence of a long-range ferroelectric order is unlikely. However, the presence of local dipoles of different magnitude could be expected under compression above 5 GPa due to the appearance of different Cu-$O_{long}$ distances, in agreement with a piezoelectric response also observed by piezoelectric current measurements (4). We confirm the appearance of a modest piezoelectric tensor induced by compression in the 5-13 GPa pressure range.

The A and B XANES features vary their central energy position with pressure. The edge region on the XANES at Cu K edge reflects dipolar electronic transitions from 1*s* to 4*p* empty levels, right above the Fermi level. In particular, the B shoulder originates both from transitions to the $4p\pi$ apical orbitals and shakedown processes through ligand to metal charge transfer. Its moderate shift is related to variations on the charge transfer energy, as probed by the calculated effective charge transfer from Cu to O ions (figure 6 (c)) and to the increase of the energy band gap.

Moreover, the discontinuity of the B structure at 5 GPa, which also reflects the transitions to energy states right above the Fermi level (*p* states), can be the signature of an electronic transition (27), also correlated with the onset of local dynamical disorder observed by EXAFS analysis at the same pressure. This might affect the macroscopic transport and dielectric properties of the material and could explain the drop in DC resistance at 3 - 4.5 GPa (4), coinciding approximately with the discontinuity in the B shoulder and the change in the $E_{gap}$ trend with pressure.

Summarizing, our XAS study of CuO under pressure revealed a transition to a dynamically distorted Cu-O local structure at 5 GPa, with a $Cu^{2+}$ octahedral-like ion (4+2) above 13 GPa. In addition, the description of the evolution of the local structure obtained from EXAFS, together with the *ab-initio* calculations, allowed us to interpret XANES A and B features and their evolution with pressure. Our results probe the correlation between the local structural, vibrational and electronic properties on CuO under pressure.

## ACKNOWLEDGMENTS

The authors thank the ESRF for granting beamtime for the proposal HC-3339 and BM23 beamline staff for experimental and excellent technical support. V. M. acknowledges the "Juan de la Cierva" fellowship (FJCI-2016-27921). A.O.R. thanks the Spanish government for a Ramón y Cajal fellowship (RyC-2016-20301) and for financial support (projects PGC2018-097520-A-100 and RED2018-102612-T), and the MALTA Consolider supercomputing centre and Compute Canada for computational resources. Premier Research Institute for Ultrahigh-pressure Sciences (PRIUS) is acknowledged for providing of nanodiamond anvil cells for XAS experiments. We would also like to acknowledge Dra. Gloria Subías and Dr. Javier Ruiz Fuertes for fruitful discussions.

## ABBREVIATIONS

XAS, X-ray absorption spectroscopy; EXAFS, extended X-ray absorption fine structure; MF, multiferroicity; DW, Debye-Waller; RT, room temperature; AFM, antiferromagnetic; BZ, Brillouin zone; IR, infra-red; CT, charge transfer.

## Supplementary Information.

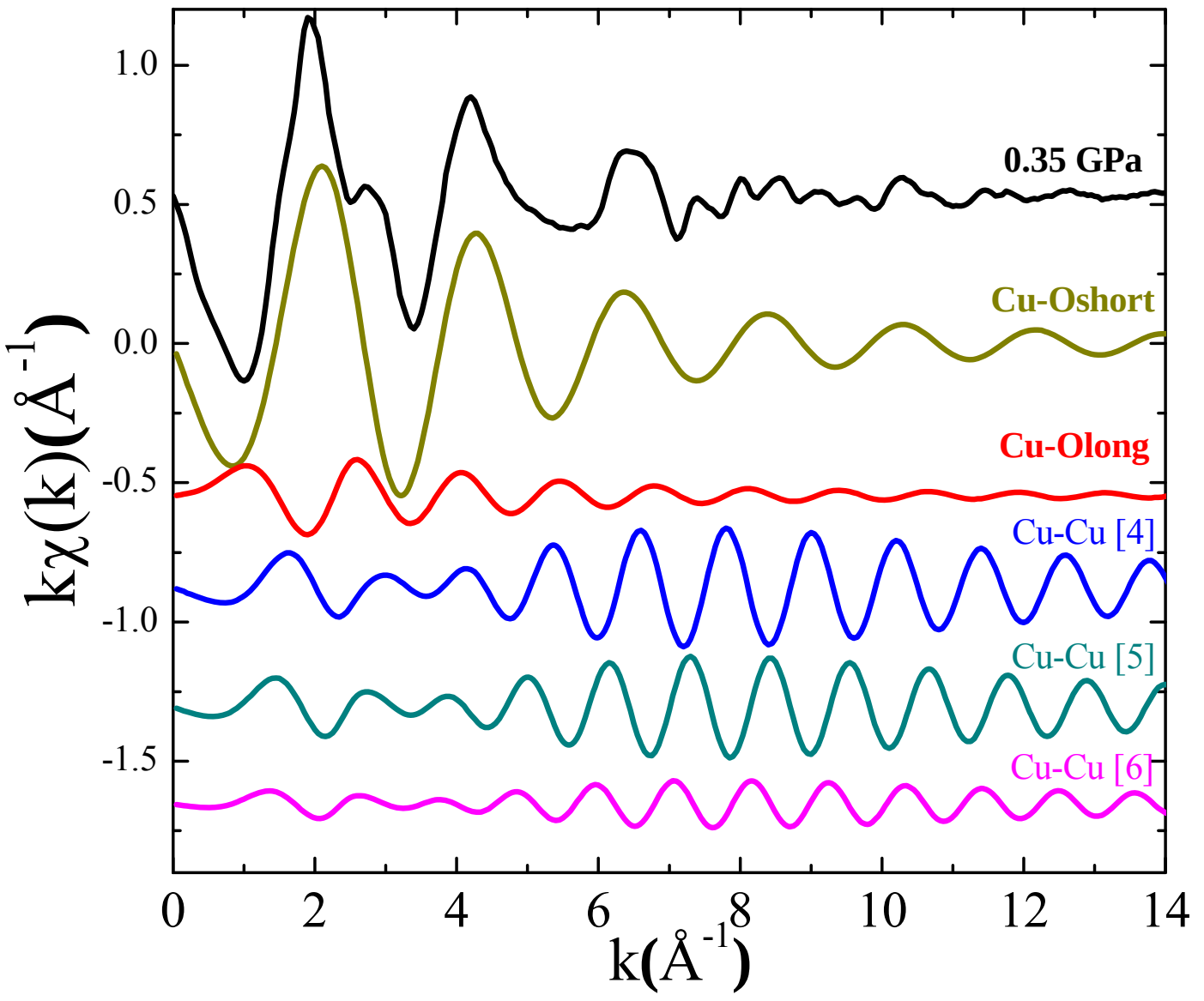

**Fig. S1.** EXAFS signal weighted in k for the reference spectra (0.35 GPa) and the calculated contribution for the different paths used in the fit, i.e. [1] Cu-O (N=2) at 1.95 Å, [2] Cu-O (N=2) at 1.96 Å, [3] Cu-O (N=2) at 2.78 Å, [4] Cu-Cu (N=4) at 2.90 Å, [5] Cu-Cu (N=4) at 3.08 Å and [6] Cu-Cu (N=2) at 3.17 Å, where [1] and [2] are jointly presented as Cu-Oshort and [3] corresponds to Cu-Olong. The spectra are shifted along y-axis for the sake of clarity.

**Table S1.** Fit structural parameters used in the model for the EXAFS data. The amplitude reduction factor is fixed to $S_0^2 = 0.78$ and $\Delta E_0 = 0.9$ eV, values found for 0.35 GPa data.

| **Path** | **R (Å)** | **delr** | $\sigma^2$**(Å$^2$)** |
|---|---|---|---|
| *Cu – O [1]* | 1.951 | delr1 | $\sigma^2$_1 |
| *Cu – O [2]* | 1.961 | delr1 | $\sigma^2$_1 |
| *Cu – O [3]* | 2.784 | delr2 | $\sigma^2$_2 |
| *Cu – Cu [4]* | 2.901 | delr3 | $\sigma^2$_3 |
| *Cu – Cu [5]* | 3.083 | delr4 | $\sigma^2$_4 |
| *Cu – Cu [6]* | 3.173 | delr4 | $\sigma^2$_4 |

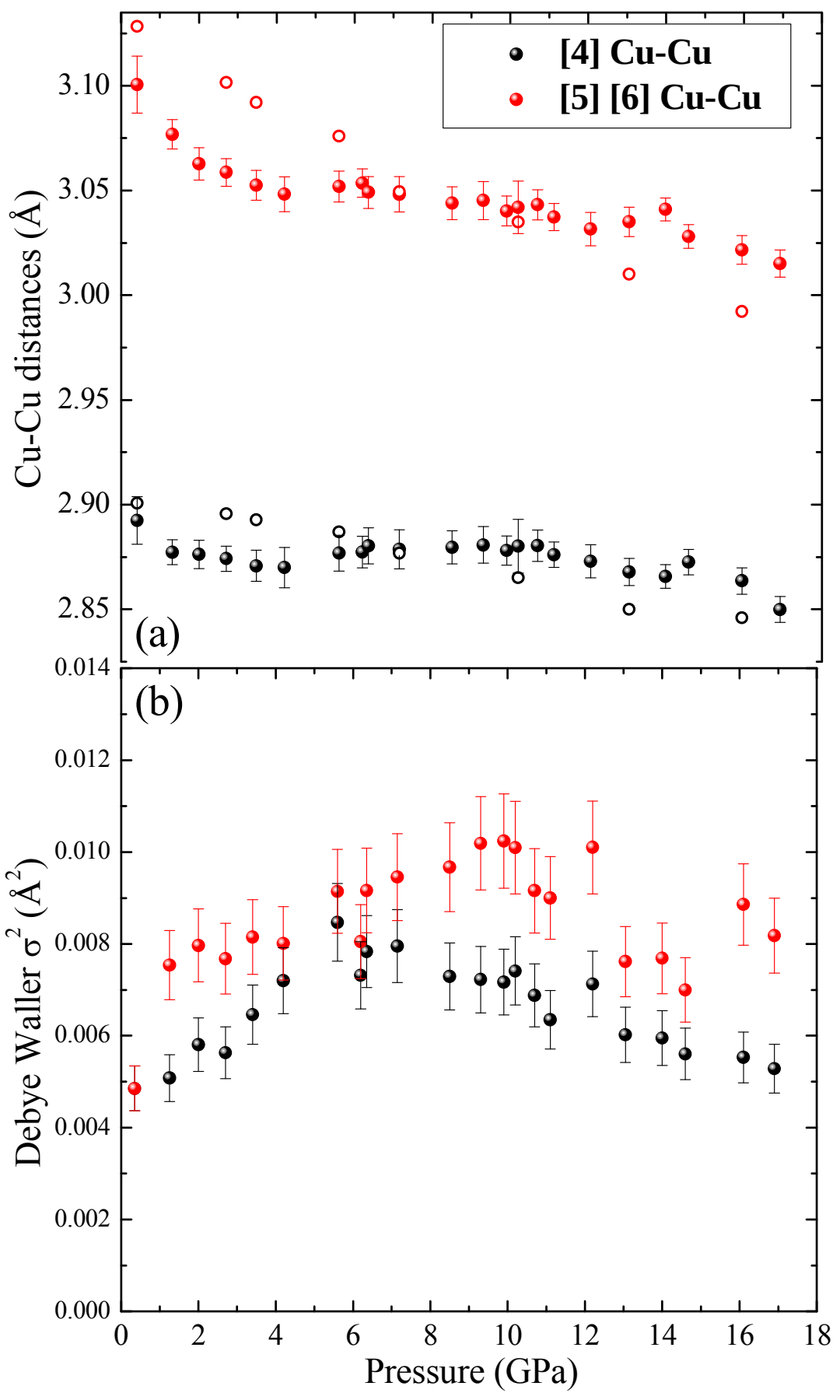

**Fig. S2.** (a) Evolution of Cu-Cu distances with pressure. In open symbols the evolution found by x-ray diffraction data taken from ref. (5) is represented. (b) Evolution of the $\sigma^2$ of the shorter (black) and longer (red) Cu-Cu distances with pressure.